# Double-well potentials and crucial estimations in nonlinear dynamics of microtubules

Rama Gupta[1], Nicolina Pop[2], Dragana Ranković[3], Slobodan Zdravković[4,5]

[1]*DAV University, Jalandhar, India*
[2]*Department of Physical Foundations of Engineering, Politehnica University of Timisoara, Timisoara, Romania*
[3]*Farmaceutski fakultet, Univerzitet u Beogradu, 11221 Beograd, Serbia*
[4]*Institut za nuklearne nauke Vinča, Univerzitet u Beogradu, Serbia*
[5]*Serbian Academy of Nonlinear Sciences*

*e-mail: szdjidji@vin.bg.ac.rs*

A B S T R A C T

In the present work, we study the two-component model of microtubules, the basic components of the eukaryotic cytoskeleton. We introduce a couple of estimations, which tremendously simplified the model. The paper is devoted to tangential oscillations of dimers, but we explain that the model can explain the radial oscillations as well. Finally, we study the stability of all solutions of differential equations, describing the dynamics of the microtubules.

*Corresponding author.
*E-mail address:* szdjidji@vin.bg.ac.rs (S. Zdravković)

## 1. Introduction

An important characteristic of eukaryotic cells is the existence of intracellular protein filament networks. Microtubules (MTs) are the basic components of this cytoskeleton. It is a long hollow, cylindrical polymer structure that spreads between the nucleus and cell membrane. We assume that its structure is known to the readers [1-5] and will be only outlined here. Its surface is formed typically by 13 long structures called protofilaments (PFs), representing a series of electric dipoles called dimers. For most of the models, a dimer is a constitutive unit, which means that its internal structure is not taken into consideration. Its mass and length are: $m = 1.8\times10^{-22}\,\mathrm{kg}$ [4, 6] and $l = 8\,\mathrm{nm}$, respectively [3, 6, 7].

There are a few models describing the nonlinear dynamics of MTs and almost all of them assume one degree of freedom per dimer. The first of them was published more than 30 years ago [6]. Its improved and more general version is so-called $u$-model [8]. Depending on the coordinate describing the dimer's displacement, each model can be seen as either longitudinal or angular. In the case of the $u$-model and its precedent, the dimers perform angular oscillations, but the choice of the coordinate $u$ places them among longitudinal models [6, 8]. There also exist angular models with the coordinate $\varphi$ being the angle between directions of the dimer and the appropriate electric field, around which the dimer oscillates [9, 10].

In this paper we deal with the two-component (2-C) model. It was introduced relatively recently [11, 12]. We used a continuum approximation and came up with solitary waves moving along MT. A different mathematical procedure, within the same model, brought about breather solitons [13].

## 2. Two component model of microtubules

Figure 1 shows a dimer that oscillates between the positions OA and OB around electric field $\vec{E}_1$. Index 1 has been introduced because there are two possible orientations for the electric field, which will be explained later on. It is obvious that we need two coordinates to describe this oscillation. The angle $\theta$ describes the position of $\vec{E}_1$, while $\varphi$ is a displacement from the direction of $\vec{E}_1$. It is obvious that

$$\theta = \theta_0 + \varphi \,. \tag{1}$$

A resultant internal electric field, $\vec{E} = \vec{E}_1 + \vec{E}_2$, coming from all dimers, is in the direction of PF. The dimer does not oscillate around the direction of $\vec{E}$ because any displacement would move it towards the directions of either $\vec{E}_1$ or $\vec{E}_2$. This will be further elaborated in Section 3. Let us point out that $\vec{E}$ is the internal electric field, unless the dimer is intentionally put into a certain external electric field for the sake of an experiment.

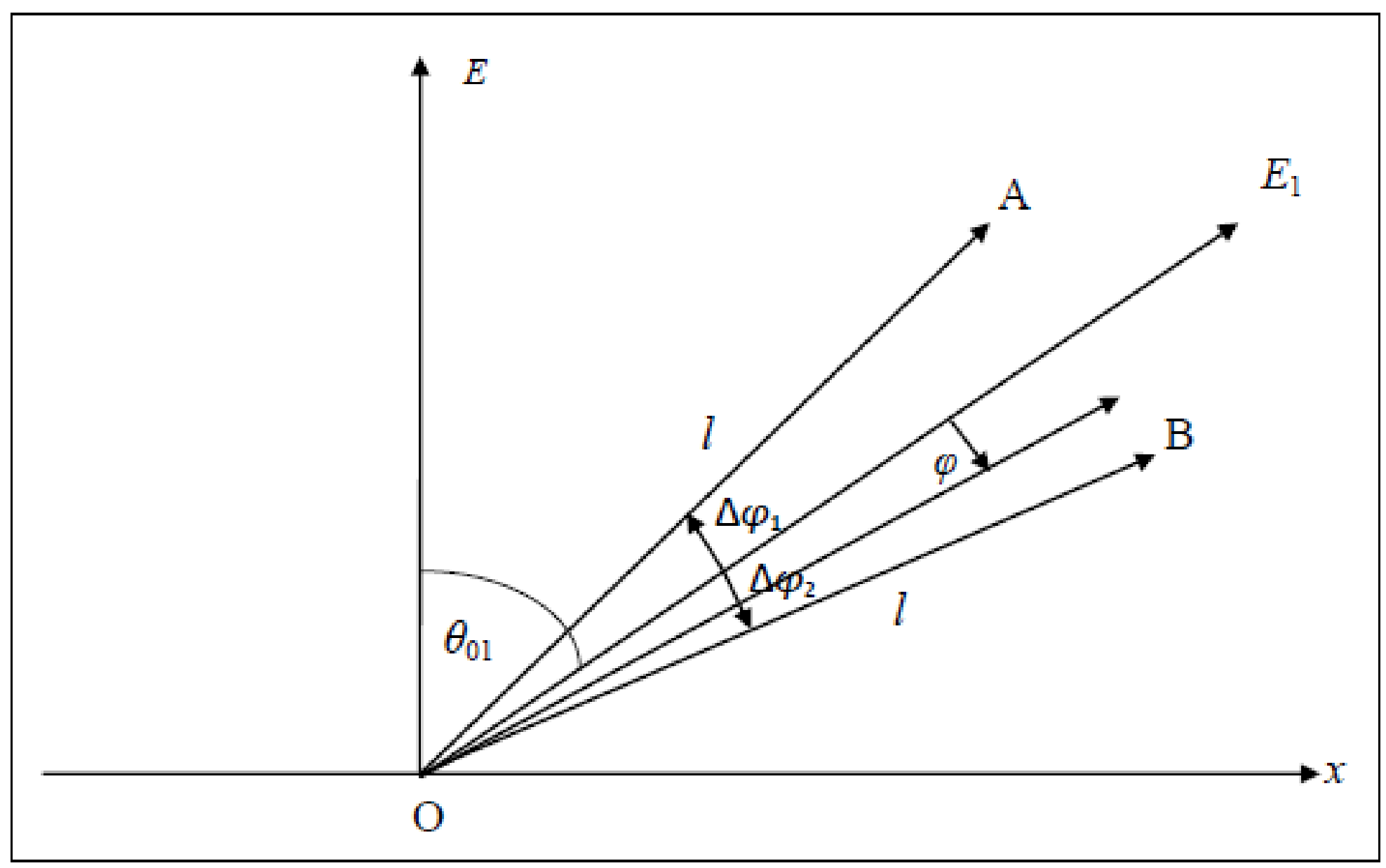

**Fig. 1.** A schematic representation of the dimer's oscillation.

The Hamiltonian for the MT can be written as

$$H = \sum_n \left[ \frac{I}{2} \dot{\varphi}_n^{\ 2} + \frac{k}{2} (\varphi_{n+1} - \varphi_n)^2 + W(\theta_n) - pE_1 \cos \varphi_n \right], \tag{2}$$

where $n$ stands for the dimer's position. The first term is the kinetic energy, while the second one is the potential energy of the chemical interaction between neighbouring dimers belonging to the same PF in the nearest-neighbour approximation. The dot means the first derivative with respect to time, $I$ is a moment of inertia of the single dimer, and $k$ is the inter-dimer stiffness parameter.

The term $W(\theta_n)$ is the well-known double-well potential energy [6]. It determines the angles $\theta_{01}$, shown in Fig. 1, and $\theta_{02}$. Hence, the overall effect of the surrounding dipoles on a chosen position $n$ can be qualitatively described by either the symmetric double-well potential

$$W_1(\theta_n) = -\frac{A}{2}\theta_n{}^2 + \frac{B}{4}\theta_n{}^4, \quad A > 0, \quad B > 0, \tag{3}$$

or the unsymmetrical one

$$W_2(\theta_n) = -\frac{A}{2}\theta_n{}^2 + \frac{B}{4}\theta_n{}^4 - \frac{D}{3}\theta_n{}^3. \tag{4}$$

The function $W_1(\theta_n)$ has two symmetric minima at $\theta_{01}$ and $-\theta_{01}$, while $W_2(\theta_n)$ has the minima at $\theta_{01}$ and $\theta_{02}$. This is shown in Fig. 2. When we exchange the sign of $D$, nothing serious happens, as only the minima $\theta_{01}$ and $\theta_{02}$ exchange the positions. Hence, we can assume $D > 0$. The case

$$W_3(\theta_n) = -\frac{A}{2}\theta_n{}^2 + \frac{B}{4}\theta_n{}^4 - C\theta_n, \qquad C > 0 \tag{5}$$

also represents an unsymmetrical potential [8, 11, 14], but is not a topic of this paper.

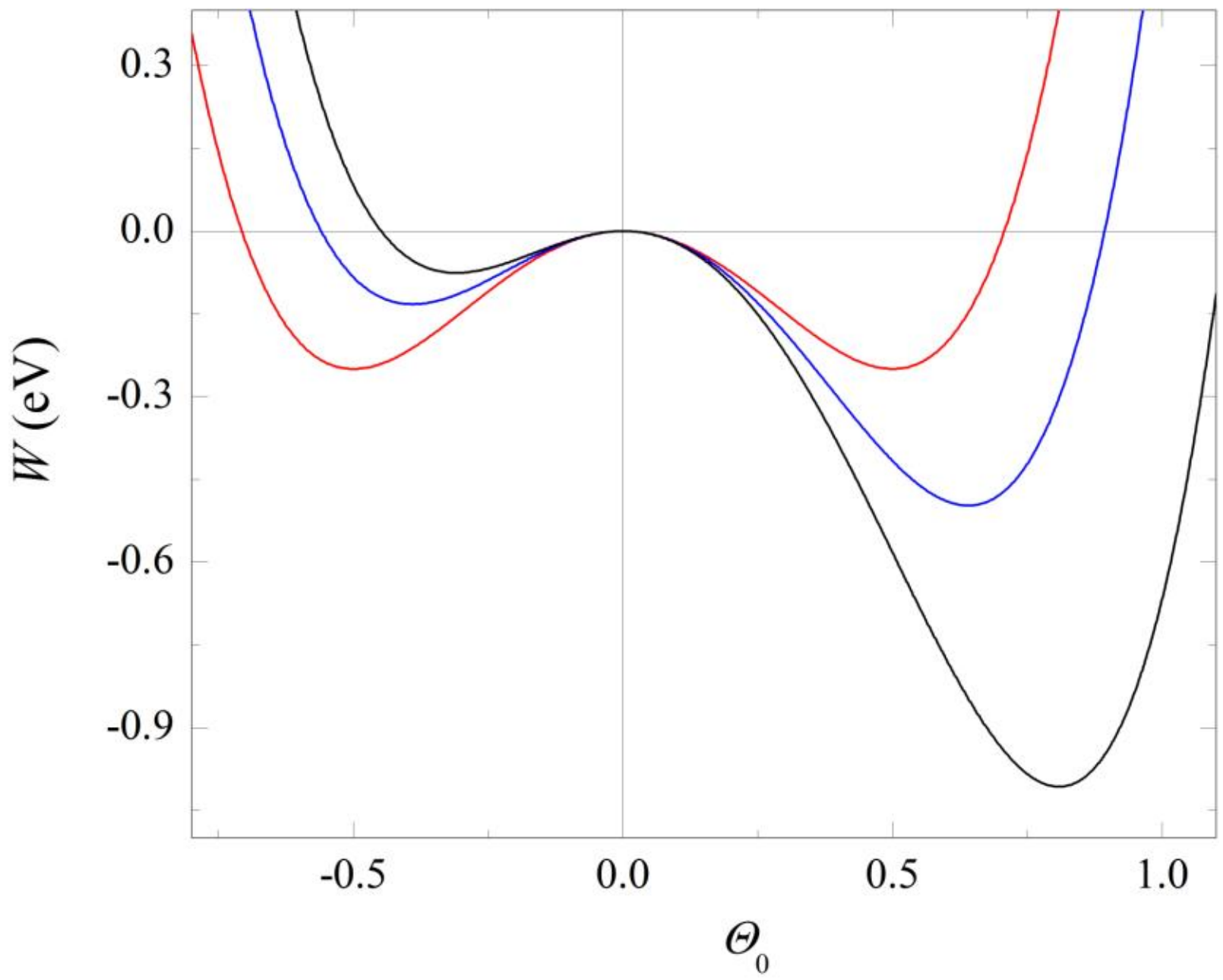

**Fig. 2.** The functions $W_1(\theta)$ (red) and $W_2(\theta)$ (blue and black) for $A = 4\text{eV}$, $B = 16\text{eV}$, $D = 4\text{eV}$ (blue), and $D = 8\text{eV}$ (black).

The last term in Eq. (2) comes from the fact that the dimer is an electric dipole which exists in the field of all other dimers. The electric field $E_1$ is our arbitrary choice explained above, while $p$ is an electric dipole moment. It is assumed that $p > 0$ and $E_1 > 0$.

In this paper we study the MT dynamics using both $W_1(\theta_n)$ and $W_2(\theta_n)$. Before we perform this job, we should do some estimation in Section 3.

**3. Estimations**

Let us concentrate on Eq. 3 and the red line in Fig. 2. One can easily show that

$$W_{1\min} = -\frac{A^2}{4B} \quad \text{at} \quad \theta_0 = \sqrt{\frac{A}{B}}. \tag{6}$$

It is known that supply of energy from hydrolysis of guanosine triphosphate, which is about $W_0 = 0.25\text{eV}$, may excite the vibrations in MTs [15]. Hence, we can assume

$$W_0 = \frac{A^2}{4B} = \frac{1}{4}\text{eV}. \tag{7}$$

To estimate the values for $\theta_{01}$, we should keep in mind the structure of MT. Fig. 3 shows one short part of the MT, that is, only three dimers in the neighbouring PFs. It is known that $l = 8\text{nm}$, $d = 4\text{nm}$ [16], $a = 0.6l = 4.8\text{nm}$, which comes from the geometry of MTs, and the width of the dimers, not indicated in the figure, is about $4\text{nm}$ [3]. Hence, the distance between the neighbouring PFs is only about $0.8\text{nm}$. The dimers in Fig. 3 are orientated in the direction of the MT. As was mentioned above, this would not be a stable position. The dimers are shifted to a certain angle. We can think that a positive side of the dimer A is attracted by the negative one in B, which contributes to the shift. Also, the interaction between A and C may bring about a different angle. These shifts are the angles $\theta_{01}$ and $\theta_{02}$. As the distance between the PFs is small, about $0.8\text{nm}$, these angles cannot be big. Let us assume that there should be $\theta_{01} \le \frac{\pi}{6}$, that is, $\sqrt{\frac{A}{B}} \le \frac{1}{2}$, which brings about

$$B \ge 4A. \tag{8}$$

Eqs. (7) and (8) yield

$$A \ge 4\text{eV}, \quad B \ge 16\text{eV}. \tag{9}$$

Obviously, we are using $1/2$ for the calculations instead of $\pi/6$, which was mentioned when we estimated the angle.

All this is based on an assumption that the dimers perform the oscillations in the tangential plane. In this case, the angles $\theta_{01}$ and $\theta_{02}$ should be small. However, they can oscillate in the radial plain. This will be discussed in Section 6.

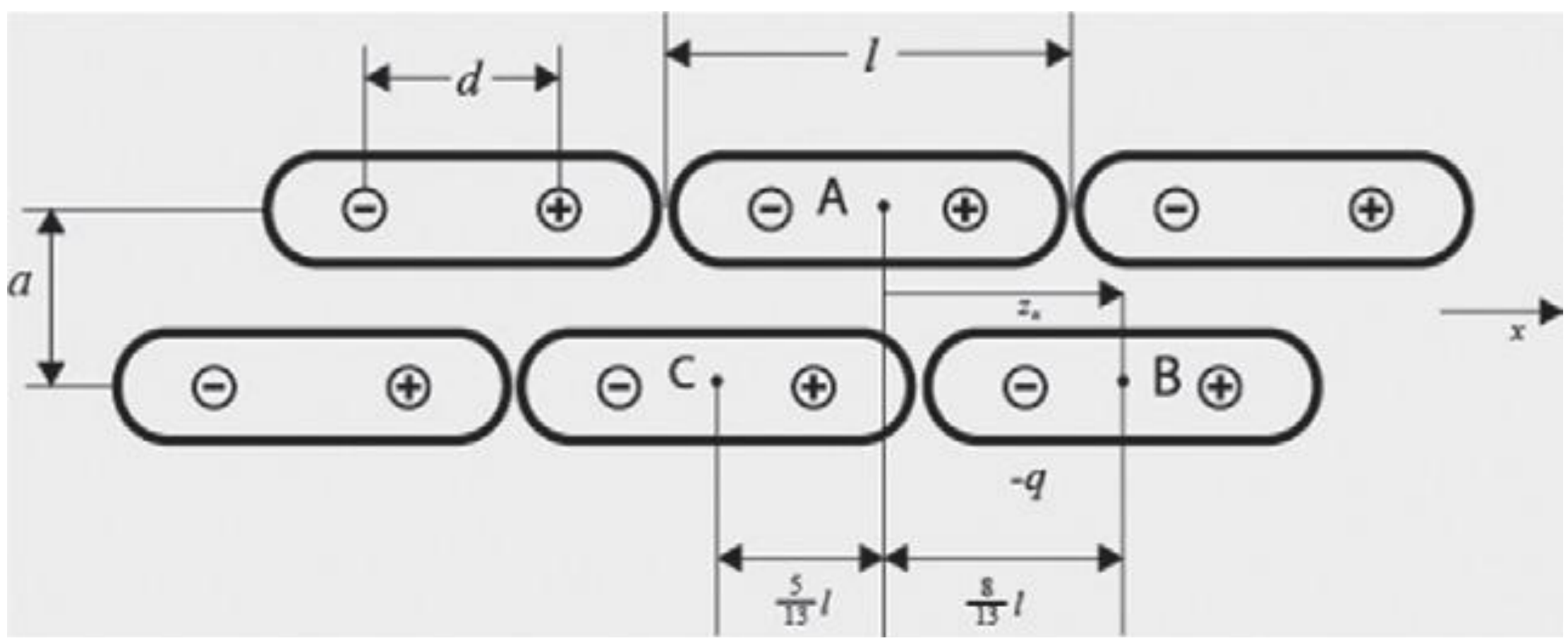

**Fig. 3.** A segment of two neighbouring protofilaments (Reproduced with a permission from Ref. [17])

Let us now estimate the product $pE$, where the index 1 was dismissed. The internal electric field in the direction of MT was estimated to be $E = 1.7\times10^{7}\,\dfrac{\mathrm{N}}{\mathrm{C}}$ [17]. For $p = 337\mathrm{D} = 1.13\times10^{-27}\mathrm{Cm}$ [3] we get

$$pE = 0.12\mathrm{eV}\,. \tag{10}$$

We will see that these estimations are very helpful. In some of the previous papers, where we were dealing with older models of MTs, we had to study many cases. Now, due to the estimated values, we deal only with the cases having physical sense.

In what follows, we study MT dynamics using the potentials $W_1(\theta_n)$ and $W_2(\theta_n)$ in Eq. (2).

**4. MT dynamics based on $W_1(\theta_n)$ in Eq. (2)**

Let us return to the Hamiltonian (2) and Eq. (3). Using Hamilton's equations of motion and a serious expansion of the cosine function, we straightforwardly get the following dynamical equation of motion:

$$I\ddot{\varphi}_n = k(\varphi_{n+1} + \varphi_{n-1} - 2\varphi_n) - A_0\varphi_n - C_0\varphi_n{}^2 - B_0\varphi_n{}^3 + D_0 \,, \tag{11}$$

where

$$A_0 = -A + 3B\theta_0{}^2 + pE \,, \quad B_0 = B - pE/6 \,, \quad C_0 = 3B\theta_0 \,, \quad D_0 = A\theta_0 - B\theta_0{}^3 \,. \tag{12}$$

A first derivative of $W_1(\theta_n)$ suggests that there should be

$$D_0 = A\theta_0 - B\theta_0{}^3 = 0 \,. \tag{13}$$

This gives the following three cases:

$$\theta_0 = \theta_0{}^{(1)} = 0 \,, \quad \theta_0 = \theta_0{}^{(2)} = \sqrt{\frac{A}{B}} \,, \quad \theta_0 = \theta_0{}^{(3)} = -\sqrt{\frac{A}{B}} \,. \tag{14}$$

These three cases give three values for each parameter in Eq. (12), that is,

$$A_0{}^{(1)} = -A + pE < 0 \,, \ A_0{}^{(2)} = A_0{}^{(3)} = 2A + pE > 0 \,, \ C_0{}^{(1)} = 0 \,, \ C_0{}^{(2)} = -C_0{}^{(3)} = 3\sqrt{AB} \,. \tag{15}$$

Of course, the parameter $B_0$, existing in Eq. (12), is not affected by the different values of $\theta_0$.

Using a continuum approximation, we get the following PDE from Eq. (11)

$$I\frac{\partial^2\varphi}{\partial t^2} - kl^2\frac{\partial^2\varphi}{\partial x^2} + A_0\varphi + C_0\varphi^2 + B_0\varphi^3 + \Gamma\frac{\partial\varphi}{\partial t} = 0 \,, \tag{16}$$

where the last term is a viscosity momentum introduced to get a more realistic equation, and $\Gamma$ is a viscosity parameter. To obtain travelling wave solutions we need a new coordinate

$$\xi = \kappa x - \omega t \,, \tag{17}$$

where $\kappa$ and $\omega$ are constants. We study all three cases, determined by Eqs. (14) and (15).

**Case 1** ( $A_0{}^{(1)} = -A + pE < 0 \,, \quad B_0 > 0 \,, \quad C_0{}^{(1)} = 0$ )

Eqs. (16) and (17) give

$$\frac{I\omega^2 - kl^2\kappa^2}{-A_0}\varphi'' - \frac{\Gamma\omega}{-A_0}\varphi' - \varphi + \frac{B_0}{-A_0}\varphi^3 = 0 \,, \tag{18}$$

which brings about

$$\alpha\psi'' - \rho\psi' - \psi + \psi^3 = 0 \tag{19}$$

for

$$\varphi = \sqrt{\frac{-A_0}{B_0}}\psi\,,\quad \alpha = \frac{I\omega^2 - kl^2\kappa^2}{-A_0}\,,\quad \rho = \frac{\Gamma\omega}{-A_0} > 0\,. \tag{20}$$

Of course, $\psi'$ and $\psi''$ are derivatives with respect to $\xi$. Here, $A_0$ stands for $A_0^{(1)}$.

Our next step is a solution of Eq. (19). There are a couple of mathematical methods for solving it, including a standard procedure [6, 18], the $(G'/G)$-expansion method [19, 20], a procedure based on Jacobian elliptic functions [21, 22], an exponential function procedure [23, 24], the method of factorisation [25-27], the simplest equation method [28, 29], and so on. As we are looking for solutions having physical sense, we here apply one of the simplest methods called the tanh-function method (THFM) [30-33]. We look for the possible solution of Eq. (19) in the form [30, 31]

$$\psi = a_0 + \sum_{i=1}^{M}\left(a_i\Phi^i + b_i\Phi^{-i}\right), \tag{21}$$

where the function $\Phi = \Phi(\xi)$ is a known solution of a differential equation of the lower order than Eq. (19). Such a differential equation may be the well-known Riccati equation, whose simplified version is

$$\Phi' = b + \Phi^2\,, \qquad b = \text{const}\,, \tag{22}$$

where $\Phi'$ is the first derivative. For negative $b$, its acceptable solution is

$$\Phi = -\sqrt{-b}\,\tanh\left(\sqrt{-b}\,\xi\right). \tag{23}$$

Other solutions of Eq. (22) are diverging ones. It is very easy to show that there should be $M = 1$ in Eq. (21) [33, 34]. According to Eqs. (21) and (23), we should assume $b_i = 0$ to get rid of the diverging solutions of Eq. (19). Hence, we should only determine the real parameters $a_0$, $a_1$, and $b$. We will see in what follows that the procedure allows us to determine the parameter $\alpha$, existing in Eq. (19). This parameter is extremely important. From Eqs. (2), (16), and (20), we conclude that its negative sign means that the elastic term is bigger than the inertial one and vice versa. Also, Eq. (20) can be written as

$$\alpha = \frac{I\omega^2 - kl^2\kappa^2}{-A_0} = \frac{I\kappa^2(\omega^2/\kappa^2 - kl^2/I)}{-A_0} \equiv \frac{I\kappa^2(v^2 - c^2)}{-A_0}\,, \tag{24}$$

where $v$ and $c$ are the solitary wave and sound velocities, respectively. This means that the sign of $\alpha$ shows if the wave is subsonic or supersonic.

To solve Eq. (19), we plug

$$\psi = a_0 + a_1\Phi\,, \qquad a_1 \neq 0\,, \tag{25}$$

into it and get the expression

$$K_3\Phi^3 + K_2\Phi^2 + K_1\Phi + K_0 = 0\,, \tag{26}$$

which is satisfied if all the coefficients $K_k$ are simultaneously equal to zero. This gives the following system of four equations:

$$\left.\begin{aligned} & a_0 - a_0{}^3 + a_1 b\rho = 0 \\ & 1 - 3a_0{}^2 - 2\alpha b = 0 \\ & 3a_0 a_1 - \rho = 0 \\ & 2\alpha = -a_1{}^2 \end{aligned}\right\}. \tag{27}$$

Using the software Mathematica, we get

$$a_0^{(1)} = -\frac{1}{2}, \quad a_1^{(1)} = -\frac{2\rho}{3}, \quad a_0^{(2)} = \frac{1}{2}, \quad a_1^{(2)} = \frac{2\rho}{3}, \quad b = -\frac{9}{16\rho^2}, \quad \alpha = -\frac{a_1{}^2}{2} \tag{28}$$

and the final solutions are:

$$\psi_1{}^{(1)} = -\frac{1}{2}\left[1 - \tanh\left(\frac{3}{4\rho}\xi\right)\right], \qquad \psi_1{}^{(2)} = -\psi_1{}^{(1)}. \tag{29}$$

The solutions $\psi_1{}^{(1)}$ and $\psi_1{}^{(2)}$ are kink and antikink solitons, respectively. This case ($\theta_0 = 0$) corresponds to the maximum of the function $W$ (red line in Fig. 2). The function $\psi_1{}^{(2)}$ is shown in Fig. 4 (blue). Notice that the negative $\alpha$ means that we deal with the subsonic solitons, which can be seen from Eq. (24), as $A_0 < 0$.

Let us mention one more interesting point. Due to our estimations, we studied only the case $A_0 B_0 < 0$. If we had not known the sign of this product, we would have also studied the case $A_0 B_0 > 0$. In this case, Eqs. (19) and (20) would be

$$\alpha\psi'' - \rho\psi' + \psi + \psi^3 = 0 \tag{30}$$

and

$$\varphi = \sqrt{\frac{A_0}{B_0}}\psi\,, \quad \alpha = \frac{I\omega^2 - kl^2\kappa^2}{A_0}\,, \quad \rho = \frac{\Gamma\omega}{A_0} > 0\,. \tag{31}$$

Following the same procedure, we would obtain imaginary $a_0$ and $a_1$, which means that this case is not acceptable. This might be considered as indirect proof that $A_0^{(1)}$, defined in Eq. (15), and $B_0$, defined in Eq. (12), have different signs. Of course, this holds for $\theta_0 = \theta_0^{(1)} = 0$ only.

**Case 2** ( $A_0^{(2)} = 2A + pE > 0$ , $B_0 > 0$ , $C_0^{(2)} = 3\sqrt{AB}$ )

In this case, both $A_0$ and $B_0$ are positive, which can be seen from Eqs. (15) and (12). Following the same procedure as in Case 1, we come up with

$$\alpha\psi'' - \rho\psi' + \psi + \delta\psi^2 + \psi^3 = 0, \tag{32}$$

where

$$\alpha = \frac{I\omega^2 - kl^2\kappa^2}{A_0}, \quad \rho = \frac{\Gamma\omega}{A_0}, \quad \delta = \frac{C_0}{\sqrt{A_0 B_0}}, \quad \varphi = \sqrt{\frac{A_0}{B_0}}\psi . \tag{33}$$

Eq. (32) appears in Refs. [11] and [12]. Hence, its solutions are [12]:

$$\psi_2^{(1)} = -\frac{\delta}{2} + \frac{K}{2}\tanh\left(\frac{\delta K}{4\rho}\xi\right), \tag{34}$$

$$\psi_2^{(2)} = -\frac{\delta - K}{4} - \frac{1}{K_0}\tanh\left(\frac{\delta + 3K}{4\rho K_0}\xi\right), \tag{35}$$

and

$$\psi_2^{(3)} = -\frac{\delta + K}{4} + \frac{K_0}{4}\tanh\left(\frac{3K - \delta}{16\rho}K_0\xi\right), \tag{36}$$

where

$$K = \sqrt{\delta^2 - 4}, \quad K_0 = K + \delta . \tag{37}$$

Also,

$$\alpha = -\frac{a_1^2}{2}, \tag{38}$$

where all three values of $a_1$ are given in Ref. [12]. The negative $\alpha$ means that we deal with the subsonic solitons again.

Before we proceed, we want to discuss the values for the parameter $\delta$. Namely, Eq. (37) states that our solutions (34)-(36) exist for $\delta \geq 2$, but we will see that the allowed interval for $\delta$ is very narrow. Eqs. (33), (12), (14), and (15) give

$$\delta = \frac{3\sqrt{AB}}{\sqrt{(2A + pE)\left(B - \frac{pE}{6}\right)}}. \tag{39}$$

Our previous estimations (9) and (10) allow us to simplify Eq. (39). A rude approximation $pE \approx 0$ brings about

$$\delta \equiv \delta_0 = \frac{3}{\sqrt{2}} = 2.12, \tag{40}$$

while a better one $(pE)^2 \approx 0$ yields to

$$\delta = \frac{3\sqrt{AB}}{\sqrt{2AB + pE\left(B - \frac{A}{3}\right)}}. \tag{41}$$

We see that the acceptable interval for $\delta$ is

$$\delta < \delta_0 = \frac{3}{\sqrt{2}} = 2.12. \tag{42}$$

One of the three values for $a_1$ is $a_1^{(3)} = \frac{4\rho}{\delta - 3K}$ [12]. However, $\delta - 3K = 0$ for $\delta = \delta_0$. This problem was solved in Ref. [11] but now we see that the problem does not exist at all, as $\delta < \delta_0$.

Let us investigate the smallest value for $\delta$. For this, we transform Eq. (41) into

$$f_0(\delta) \equiv \frac{9}{\delta^2} - 2 = pE\left(\frac{1}{A} - \frac{1}{3B}\right). \tag{43}$$

As,

$$f_0(\delta) < \frac{pE}{A}, \tag{44}$$

we conclude, using Eq. (9), that the final interval for $\delta$ is

$$2.106 < \delta < \delta_0 = \frac{3}{\sqrt{2}} = 2.12. \tag{45}$$

Very often, while modelling physical systems, we have problems with the parameters. We must estimate or, even, guess them. In this case, the allowed interval for $\delta$ is so narrow that we almost know its exact value.

Now, we proceed with Eqs. (34)-(36). They can be simplified using Eq. (37). These more elegant solutions are

$$\left.\begin{aligned}\psi_2{}^{(1)} &= -\frac{1}{2}\left[\delta - K\tanh\left(\frac{\delta K}{4\rho}\xi\right)\right]\\ \psi_2{}^{(2)} &= -\frac{\delta - K}{4}\left[1-\tanh\left(\frac{\Delta_-}{8\rho}\xi\right)\right]\\ \psi_2{}^{(3)} &= -\frac{\delta + K}{4}\left[1-\tanh\left(\frac{\Delta_+}{8\rho}\xi\right)\right]\end{aligned}\right\}, \tag{46}$$

where

$$\Delta_\pm = \delta^2 \pm \delta K - 6. \tag{47}$$

Notice negative values for $\Delta_\pm$. For example, $\Delta_+(2.1) = -0.25$ and $\Delta_-(2.1) = -2.9$.

**Case 3** ( $A_0{}^{(3)} = 2A + pE > 0$, $B_0 > 0$, $C_0{}^{(3)} = -3\sqrt{AB}$ )

Following the same procedure as above, we get Eqs. (32) and (33) again. As $C_0{}^{(3)} < 0$, the parameter $\delta$ is negative. An alternative approach could be

$$\alpha\psi'' - \rho\psi' + \psi - \delta\psi^2 + \psi^3 = 0, \tag{48}$$

where

$$\delta = \frac{\left|C_0{}^{(3)}\right|}{\sqrt{A_0 B_0}} = \frac{3\sqrt{AB}}{\sqrt{A_0 B_0}} > 0. \tag{49}$$

The solutions of Eq. (48) can be obtained from Eq. (46) by a replacement $\delta \to -\delta$. They are:

$$\psi_3{}^{(1)} = -\psi_2{}^{(1)}, \quad \psi_3{}^{(2)} = -\psi_2{}^{(3)}, \quad \psi_3{}^{(3)} = -\psi_2{}^{(2)}. \tag{50}$$

These results are expected because when we change a sign of $\psi$ in Eq. (32) we get Eq. (48).

The functions $\psi_3{}^{(1)}$, $\psi_3{}^{(2)}$, $\psi_3{}^{(3)}$, and $\psi_1{}^{(2)}$ are shown in Fig. 4. A value $\delta = 2.11$ was used for $\psi_3{}^{(1)}$, $\psi_3{}^{(2)}$, and $\psi_3{}^{(3)}$.

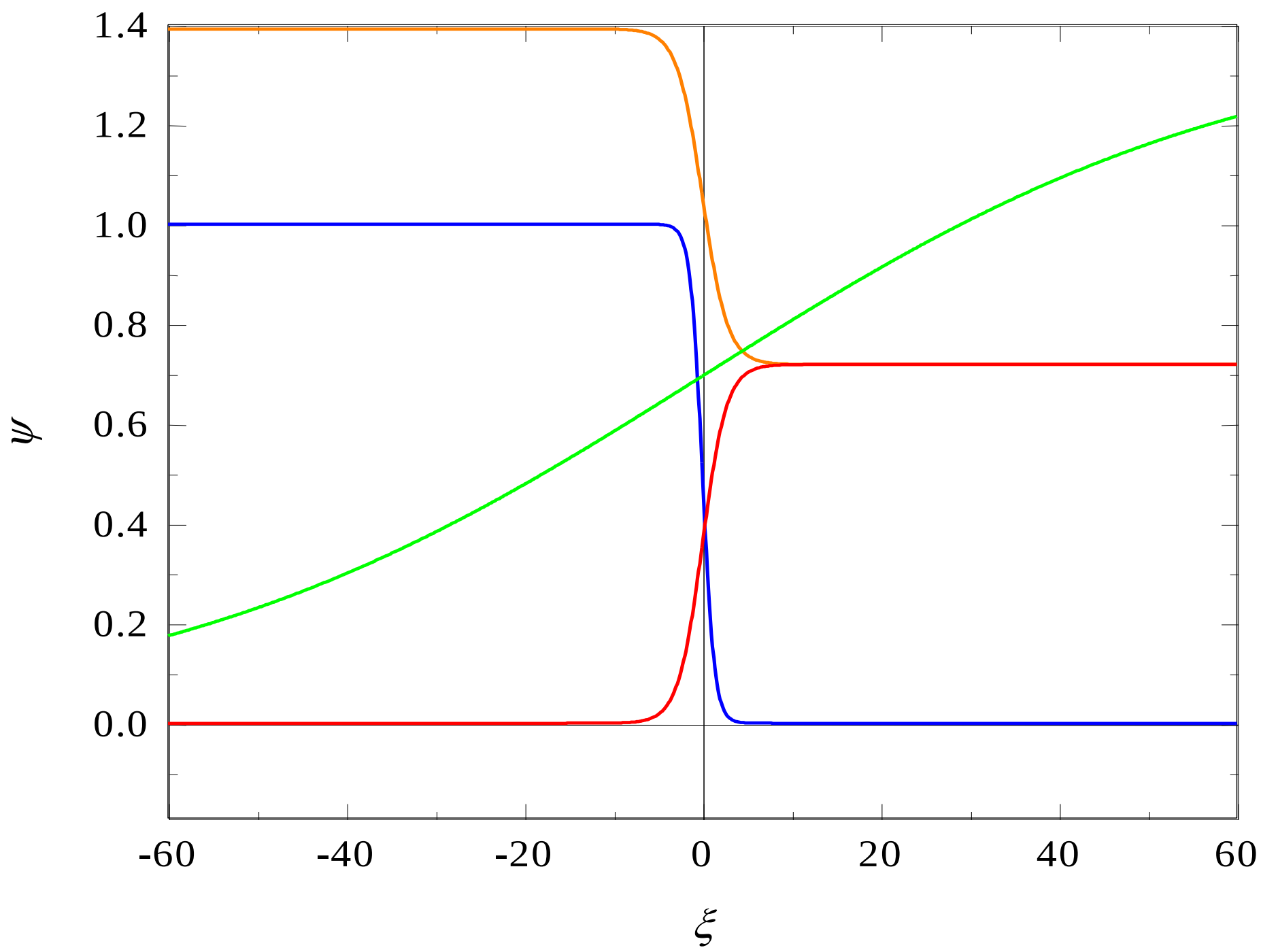

**Fig. 4.** The solutions $\psi_1^{(2)}$ (blue), $\psi_3^{(1)}$ (orange), $\psi_3^{(2)}$ (green), and $\psi_3^{(3)}$ (red) as functions of $\xi$.

## 5. Stability of all the solutions coming from $W_1(\theta_n)$

To examine the stability of the solution $\psi_1^{(1)}(\xi)$, given by Eq. (29), we perform the standard procedure [12]. This means that we introduce the substitution

$$\psi = \psi_0 + f \,, \qquad \psi_0 \equiv \psi_1^{(1)} \tag{51}$$

into Eq. (19). Denoting $\mathrm{d}f/\mathrm{d}\xi = F$ and keeping only linear terms as they determine the stability according to Lyapunov's first approximation theorem, we get the following system of the first-order differential equations:

$$\left.\begin{aligned} \frac{\mathrm{d}f}{\mathrm{d}\xi} &= F \equiv a_1 f + b_1 F \\ \frac{\mathrm{d}F}{\mathrm{d}\xi} &= \frac{\rho}{\alpha} F + \frac{1}{4\alpha} f \equiv a_2 f + b_2 F \end{aligned}\right\}. \tag{52}$$

The corresponding eigenvalues can be determined from

$$\begin{vmatrix} a_1-\lambda & b_1 \\ a_2 & b_2-\lambda \end{vmatrix} = \begin{vmatrix} -\lambda & 1 \\ \dfrac{1}{4\alpha} & \dfrac{\rho}{\alpha}-\lambda \end{vmatrix} = 0, \tag{53}$$

and they are

$$\lambda_{\pm} = \frac{-\rho \pm \sqrt{\rho^2+\alpha}}{-2\alpha}. \tag{54}$$

A solution is stable if both $\mathrm{Re}\,(\lambda_+)<0$ and $\mathrm{Re}\,(\lambda_-)<0$. The value $\lambda=0$ indicates that we cannot neglect all nonlinear terms as we did in the derivation of Eq. (52). In other words, the terms comprising $f^2$ are relevant [35]. Here, keeping in mind that $\rho>0$ and $\alpha<0$ for $\psi_0 \equiv \psi_1{}^{(1)}$, we conclude that $\psi_1{}^{(1)}$ is a stable solution. We can repeat the procedure for $\psi_0 \equiv \psi_1{}^{(2)}$ and very easily convince ourselves that this is also the stable solution.

We perform the same procedure for $\psi_0 \equiv \psi_2{}^{(1,2,3)}$. For example, for $\psi_0 \equiv \psi_2{}^{(1)}$ we get

$$\lambda_{\pm} = \frac{-\rho \pm \sqrt{\rho^2+K^2\alpha}}{-2\alpha}. \tag{55}$$

As $K^2>0$, because $\delta>2$, we conclude that $\psi_2{}^{(1)}$ is also stable.

Using the same procedure for $\psi_2{}^{(2)}$, we get

$$\lambda_{\pm} = \frac{-\rho \pm \sqrt{\rho^2+\phi_0\alpha}}{-2\alpha}, \tag{56}$$

where

$$\phi_0 = \frac{(\delta-K)^2}{4}, \qquad \alpha<0. \tag{57}$$

As $\phi_0\alpha$ is negative we conclude that $\psi_2{}^{(2)}$ is the stabile solution.

Finally, we need to investigate the stability of $\psi_2^{(3)}$. Following, again, the procedure from above, we get

$$\lambda_{\pm} = \frac{-\rho \pm \sqrt{\rho^2 + \phi_1 \alpha}}{-2\alpha}, \tag{58}$$

where

$$\phi_1 = \frac{(\delta + K)^2}{4}, \qquad \alpha < 0, \tag{59}$$

which means that $\psi_2^{(3)}$ is also the stabile solution.

Finally, we should investigate stability of the solutions $-\psi_2^{(i)}$, $i = 1, 2, 3$, of Eq. (48). According to Eq. (46), we can write $\psi_2^{(i)} = a_i + b_i \tanh(c_i)$. We plug Eq. (44) into Eqs. (32) and (48). The relevant linear terms, coming from the last three terms in Eqs. (32) and (48), are

$$T_1 = 2\delta a_i f + 3a_i^2 f \tag{60}$$

and

$$T_2 = -2\delta(-a_i) f + 3(-a_i)^2 f, \tag{61}$$

respectively. As $T_1 = T_2$, we conclude that the functions $\psi_2^{(i)}$ and $-\psi_2^{(i)}$ have the same stability, that is, all of them are stable.

## 6. MT dynamics based on $W_2(\theta_n)$ in Eq. (2)

In this section, we introduce the function $W_2(\theta_n)$ into Eq. (2). A complete analysis of this issue will be done in a more elaborated paper. The most important part of this paper is estimations of the parameters. Hence, we want to complete this research, and this is a topic of this section. We realised that the estimations are important, allowing us to study only relevant cases. Hence, we deal with the estimations again, related to $W_2(\theta_n)$.

The first derivative of $W_2(\theta_n)$, given by Eq. (4), is equal to zero for

$$\theta_0 = \theta_{01} = 0, \quad \theta_0 = \theta_{02} = \frac{D + \sqrt{D^2 + 4AB}}{2B}, \quad \theta_0 = \theta_{03} = \frac{D - \sqrt{D^2 + 4AB}}{2B}. \tag{62}$$

A positive $D$ is assumed, as explained above. We are using a notation $\theta_{0i}$ instead of $\theta_0^{(i)}$, used in Section 4, to eliminate a possible confusion. It is obvious that $\theta_{02} > 0$, $\theta_{03} < 0$, and $\theta_{02} + \theta_{03} = D/B$. Also, $W_2$ has a maximum at $\theta_{01} = 0$, that is $W_{2\max} = W_2(0) = 0$. This is shown in Fig. 2 (blue and black lines). All three lines in Fig. 2 were plotted for equal $A$ and $B$, that is, for $A = 4\text{eV}$ and $B = 16\text{eV}$. We notice that the impact of the parameter $D$ is a deeper right minimum, a shallower the left one, and a higher $\theta_{01}$. Let us keep the equal constraints as in the previous case, that is,

$$\theta_{02} = \frac{1}{2}, \qquad W_{2\min} = W_2(\theta_{02}) = -\frac{1}{4}\text{eV}. \tag{63}$$

These constraints respectively give

$$D = \frac{B}{2} - 2A, \qquad D = \frac{3B}{8} - 3A + 6. \tag{64}$$

If we eliminate $D$ from Eqs. (64), we get

$$B = 8(6 - A), \tag{65}$$

while the relationship between $A$ and $D$ is

$$D = 6(4 - A). \tag{66}$$

Hence, for $A = 4\text{eV}$, we obtain the symmetrical case as $D = 0$. In such a case, Eq. (65) gives $B = 16\text{eV}$, that is, Eq. (9). An interesting case is $A = 0$. An investigation of Eq. (4), and keeping in mind Eqs. (65) and (66), we see that the function $W_2(\theta)$ has only one minimum at $\theta_0 = \frac{D}{B} = \frac{1}{2}$ and an inflection point at $\theta_0 = 0$. This is shown in Fig. 5 (green) as well as two more examples for $A = 3.5\text{eV}$ and $A = 3\text{eV}$.

Notice that $D < 0$ for $A > 4\text{eV}$. This means that only $A \le 4\text{eV}$ makes sense as $D > 0$, which was explained above.

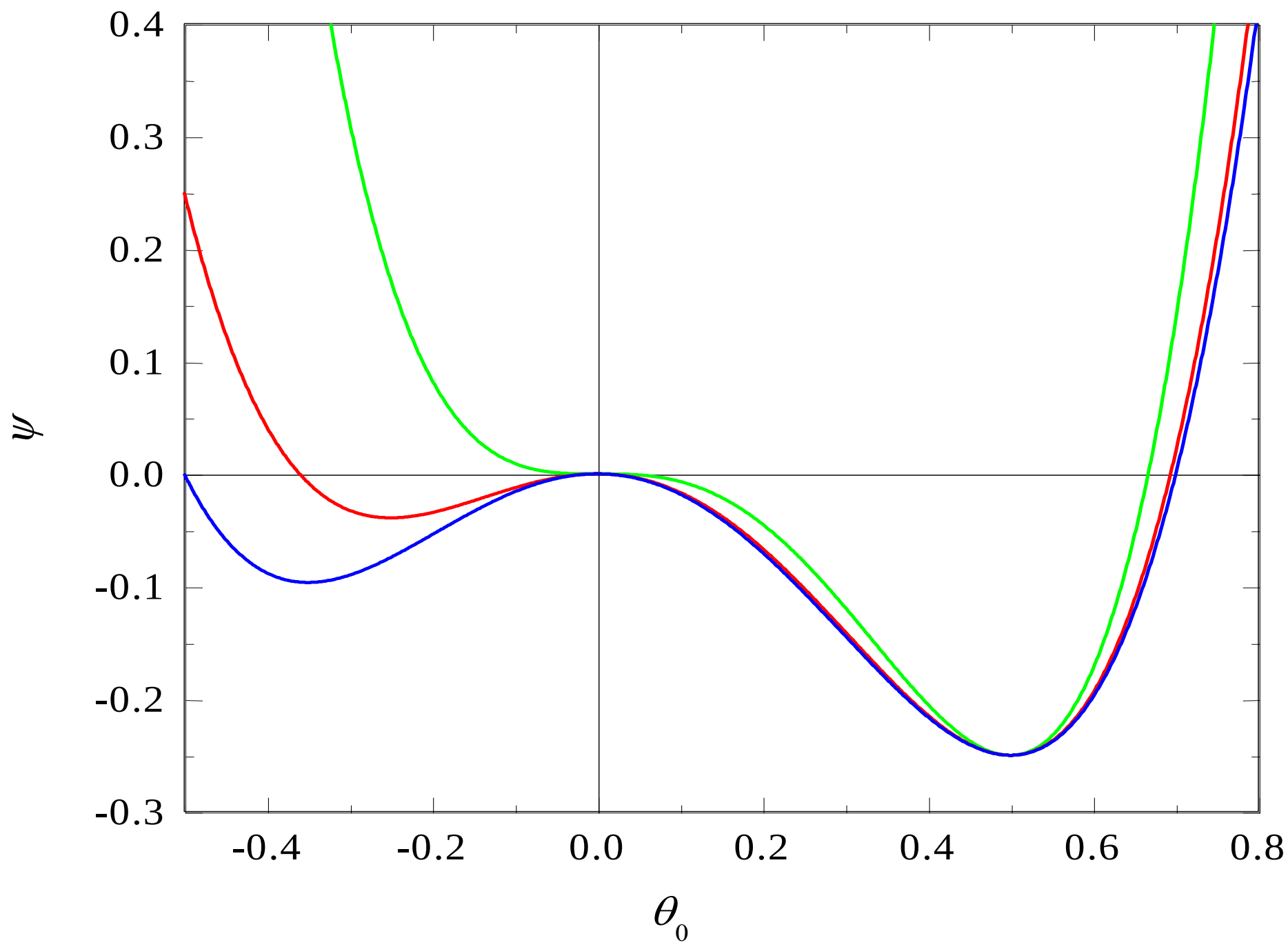

**Fig. 5.** The function $W_2$ as a function of $\theta_0$ for $A = 3.5\text{eV}$ (blue), $A = 3\text{eV}$ (red), and $A = 0$ (green).

Before we proceed, we want to discuss two more points. According to Eqs. (9) and (10), we have concluded that $A > pE$ and $B > \frac{pE}{6}$. However, we have just mentioned very small $A$. Hence, the critical values for $A = pE = 0.12\text{eV}$ are $B = 47.04\text{eV}$ and $D = 23.28\text{eV}$.

The case when the product $(A - pE)\left(B - \frac{pE}{6}\right)$ is negative may have sense in the case of the appropriate experiment. We can put MT in a strong electric field, which is in the direction of $\vec{E}$, and provide that $pE > A$. Therefore, the theory we are dealing with now can explain this situation, but this is not the topic of this paper.

Now, we return to the Hamiltonian (2) and Eq. (4). We follow the procedure explained in Section 4. Hence, Eq. (11) holds again, but the values of the parameters existing in it are:

$$A_0 = -A + 3B\theta_0^{\ 2} - 2D\theta_0 + pE, \quad B_0 = B - pE/6, \quad C_0 = 3B\theta_0 - D,$$
$$D_0 = A\theta_0 - B\theta_0^{\ 3} + D\theta_0^{\ 2}. \tag{67}$$

From $D_0 = 0$ we get the expressions for $\theta_{01}$, $\theta_{02}$, and $\theta_{03}$, that is, Eq. (62). Let us determine $A_0$ for $\theta_{02}$ and $\theta_{03}$. We eliminate $B\theta_0^{\ 2}$ and $A$ from Eq. (67) and $D_0 = 0$ and get two important equations that obviously indicate the signs of $A_0$ :

$$A_{01} = -A + pE < 0\,, \quad A_{02} = 2A + D\theta_{02} + pE > 0\,, \quad A_{03} = 2B\theta_{03}^{\ 2} - D\theta_{03} + pE > 0\,. \qquad (68)$$

as $\theta_{02} > 0$ and $\theta_{03} < 0$.

According to Eq. (67), the parameter $B_0$ does not depend on $\theta_0$ and

$$B_{01} = B_{02} = B_{03} = B - pE/6 > 0. \qquad (69)$$

Also, Eqs. (67) and (62) straightforwardly bring about

$$C_{01} = -D < 0\,, \quad C_{02} = \frac{D + 3\sqrt{D^2 + 4AB}}{2} > 0\,, \quad C_{03} = \frac{D - 3\sqrt{D^2 + 4AB}}{2} < 0\,, \qquad (70)$$

as $D > 0$.

Two possibilities are relevant here. They are $A_0 B_0 < 0$ and $A_0 B_0 > 0$. This yield to Eqs. (32) and (48). Of course, the values of the parameters are different for the cases determined by $W_1(\theta_n)$ and $W_2(\theta_n)$. The solutions of these two cases should be compared, and this will be a topic of the more elaborated paper.

## 7. Conclusions

The present paper is based on the 2-C model of MTs, which is, we believe, the most reliable in this moment. We studied two double-well potentials in the 2-C model. We should mention an advantage of $W_2$ over $W_1$. First of all, the potential $W_1$ is symmetric, which does not describe MT precisely enough, as was discussed above. Also, $W_2$ includes $W_1$ ($D = 0$) and can describe the potential energy with only one potential well ($A = 0$).

It was mentioned above, in Section 3, that we assume the tangential oscillations of the dimers, and Fig. 3 obviously suggests two minima of the potential $W_2(\theta)$. However, oscillations in the radial direction might also be possible. In this case, only one minimum of $W_2(\theta)$ is expected; the inclination of the field around which the dimer oscillates would probably be towards the outside of the MT, and the angle $\theta_0$ should not be small any more. Such a situation is shown in Ref. [36] (Fig. 1B). Therefore, the unsymmetrical double-well potential $W_2(\theta_n)$ can explain this situation, that is, the oscillation in the radial direction. However, this does not mean that the model used here is always acceptable. Namely, for the radial oscillations the angle $\theta_0$ can be big, and the series expansion of the cosine term in Eq. (2) would not be appropriate, which is an interesting challenge for future research.

**Acknowledgements**

This work was supported by a STSM Grant from COST Action CA21169, supported by COST (European Cooperation in Science and Technology). D. Ranković acknowledges that her research was funded by the Ministry of Science, Technological Development, and Innovation of the Republic of Serbia through two grant agreements with the University of Belgrade – Faculty of Pharmacy (Nos. 451-03-33/2026-03/200161 and 451-03-34/2026-03/200161).